\documentclass[12pt,a4paper]{article}

\usepackage[left=2.5cm,top=2.5cm,bottom=2.5cm,right=2.5cm]{geometry}
\usepackage{graphicx}
\usepackage{amssymb}
\usepackage{amsthm}
\usepackage{amsmath}
\usepackage{amsfonts}
\usepackage{algorithm, algorithmic}
\usepackage{color}

\usepackage{cite}

\usepackage{enumitem}

\newtheorem{thm}{Theorem}[section]

\newtheorem{lem}[thm]{Lemma}

\theoremstyle{definition}

\theoremstyle{remark}
\newtheorem{rem}{Remark}[section]

\numberwithin{equation}{section}

\begin{document}

\title{In-place associative permutation sort}
\author{A. Emre CETIN}

\maketitle

\begin{abstract}

{\em In-place associative integer sorting} technique was developed, improved and specialized for distinct integers. The technique is suitable for {\em integer sorting}. Hence, given a {\em list} $S$ of $n$ {\em integers} $S[0 \ldots n-1]$, the technique sorts the integers in ascending or descending order. It replaces bucket sort, distribution counting sort and address calculation sort family of algorithms and requires only constant amount of additional memory for storing counters and indices beside the input list. The technique was inspired from one of the ordinal theories of ``serial order in behavior" and explained by the analogy with the three main stages in the formation and retrieval of memory in cognitive neuroscience: (i) {\em practicing}, (ii) {\em storing} and (iii) {\em retrieval}. 

In this study {\em in-place associative permutation} technique is introduced for {\em integer key sorting} problem. Given a list $S$ of $n$ {\em elements} $S[0 \ldots n-1]$ each have an  {\em integer key} in the range $[0,m-1]$, the technique sorts the elements according to their integer keys in $\mathcal{O}(n)$ time using only $\mathcal{O}(1)$ amount of memory if $m<=n$. On the other hand, if $m>n$, it sorts in $\mathcal{O}(n+m)$ time for the worst, $\mathcal{O}(m)$ time for the average (uniformly distributed keys) and $\mathcal{O}(n)$ time for the best case using $\mathcal{O}(1)$ extra space.

\end{abstract}

\begin{description}
\item{keywords:} associative sort, permutation sort, stimulation sort, linear time sorting.
\end{description}

\section{Introduction}\label{sec:intro}

Nervous system is considered to be closely related and described with the ``serial order in behavior" in cognitive neuroscience~\cite{Lashley,Lashley_1} with three basic theories which cover almost all {\em abstract data types} used in computer science. These are chaining theory, positional theory and ordinal theory~\cite{Henson}.

Chaining theory is the extension of stimulus-response (reflex chain) theory, where each response can become the stimulus for the next. From an information processing perspective, comparison based sorting algorithms that sort the lists by making a series of decisions relying on comparing keys can be classified under chaining theory. Each comparison becomes the stimulus for the next. Hence, keys themselves are associated with each other. Some important examples are quick sort~\cite{Hoare}, shell sort~\cite{Shell}, merge sort~\cite{Burnetas} and heap sort~\cite{Williams}.

Positional theory assumes order is stored by associating each element with its position in the sequence. The order is retrieved by using each position to cue its associated element. This is the method by which conventional (Von  Neumann) computers store and retrieve order, through routines accessing separate addresses in memory. Content-based sorting algorithms where decisions rely on the contents of the keys can be classified under this theory. Each key is associated with a position depending on its content. Some important examples are distribution counting sort~\cite{Seward,Feurzig}, address calculation sort~\cite{Isaac,Tarter,Flores,Jones,Gupta,Suraweera}, bucket sort\cite{mahmoud:2000, Cormen} and radix sort~\cite{knuth:vol3,mahmoud:2000,sedgewick:algorithms_in_C, Cormen}.

Ordinal theory assumes order is stored along a single dimension, where that order is defined by relative rather than absolute values on that dimension. Order can be retrieved by moving along the dimension in one or the other direction. This theory need not assume either the item-item nor position-item associations of chaining and positional theories respectively\cite{Henson}.

One of the ordinal theories of serial order in behavior is that of Shiffrin and Cook\cite{Shiffrin} which suggests a model for short-term forgetting of item and order information of the brain. It assumes associations between elements and a ``node'', but only the nodes are associated with one another. By moving inwards from nodes representing the start and end of the sequence, the associations between nodes allow the order of items to be reconstructed~\cite{Henson}.

As in the ordinal model of Shiffrin and Cook, in-place associative integer sorting technique~\cite{ecetin,ecetin1,ecetin2,ecetin3} assumes that associations are between the integers in the list space and the nodes in an imaginary linear subspace that spans a predefined range of integers. The imaginary subspace can be defined anywhere on the list space $S[0\ldots n-1]$ provided that its boundaries do not cross over that of the list making the technique in-place, i.e., beside the input list, only a constant amount of memory locations are used for storing counters and indices. Hence, moving through nodes that represent the start and end of practiced integers as well as retaining their relative associations with each other even when their positions are altered by cuing allow the order of integers to be constructed in-place in linear time.

%The range of the integers spanned by the imaginary subspace is upper bounded by the number of integers $n$ but may be smaller and can be located anywhere making the technique in-place, i.e., beside the input list, only a constant amount of memory locations are used for storing counters and indices. Furthermore, this definition reveals the asymptotic power of the technique with increasing $n$ with respect to the range of integers, as well. 

Another ordinal theory is the original perturbation model of Estes~\cite{Estes}. Although proposed to provide a reasonable qualitative fit of the forgetting dynamics of the short term memory~\cite{Henson} in cognitive neuroscience, the idea behind the method is that the order of the elements is inherent in the cyclic reactivation of the elements, i.e., reactivations lead to reordering of the elements.

When the idea behind the perturbation model is combined with the original technique of associative sorting, in-place associative permutation technique is obtained where the order of the practiced interval is inherent in the cyclic reactivation of the elements of the list. Practicing phase of associative sorting is revised and when the elements of the list are reactivated with a special form of cycle leader permutation, a temporal state is obtained that can be either {\em stored} in short (or long) term memory or {\em restored} into sorted permutation of the practiced interval in linear time using $\mathcal{O}(1)$ amount of memory. Therefore, in-place associative permutation sort technique is obtained that consists of three phases namely (i) {\em practicing}, (ii) {\em permutation} and (iii) {\em restoring}. 

\subsection{Original Technique: In-place Associative Integer Sorting}

The main difficulties of all distributive sorting algorithms is that, when the integers are distributed using a hash function according to their content, several integers may be clustered around a loci, and several may be mapped to the same location. These problems are solved by inherent three basic steps of associative sort~\cite{ecetin} namely (i) {\em practicing}, (ii) {\em storing} and (iii) {\em retrieval}. 

It is assumed that associations are between the integers in the list space and the nodes in an imaginary linear subspace (ILS) that spans a predefined range of integers. The ILS can be defined anywhere on the list space $S[0\ldots n-1]$ provided that its boundaries do not cross over that of the list. The range of the integers spanned by the ILS is upper bounded by the number of integers $n$ but may be smaller and can be located anywhere making the technique in-place, i.e., beside the input list, only a constant amount of memory locations are used for storing counters and indices. An association between an integer and the ILS is created by a node using a monotone bijective hash function that maps the integers in the predefined interval to the ILS. The process of creating a node by mapping a distinct integer to the ILS is ``practicing a distinct integer of an interval''. Once a node is created, the redundancy due to the association between the integer and the position of the node releases the word allocated to the integer in the physical memory except for one bit which tags the word as a node for interrogation purposes. The tag bit discriminates the word as node and the position of the node lets the integer be retrieved back from the ILS using the inverse hash function. This is ``integer retrieval". All the bits of the node except the tag bit can be cleared and used to encode any information. Hence, they are the ``record'' of the node and the information encoded into a record is the ``cue'' by which cognitive neuro-scientists describe the way that the brain recalls the successive items in an order during retrieval. For instance, it will be foreknown from the tag bit that a node has already been created while another occurrence of that particular integer is being practiced providing the opportunity to count other occurrences. The process of counting other occurrences of a particular integer is ``practicing all the integers of an interval'', i.e., rehearsing used by cognitive neuro-scientists to describe the way that the brain manipulates the sequence before storing in a short (or long) term memory. Practicing does not need to alter the value of other occurrences. Only the first occurrence is altered while being practiced from where a node is created. All other occurrences of that particular integer remain in the list space but become meaningless. Hence they are ``idle integers''. On the other hand, practicing does not need to alter the position of idle integers as well, unless another distinct integer creates a node exactly at the position of an idle integer while being practiced. In such a case, the idle integer is moved to the former position of the integer that creates the new node. This makes associative sort unstable, i.e., equal integers may not retain their original relative order. 

Once all the integers in the predefined interval are practiced, the nodes dispersed in the ILS are clustered in a systematic way closing the distance between them to a direction retaining their relative order. This is the {\em storing} phase of associative sort where the received, processed and combined information to construct the sorted permutation of the practiced interval is stored in the short-term memory. When the nodes are moved towards a direction, it is not possible to retain the association between the ILS and list space. However, the record of a node can be further used to encode the absolute (former) position of that node as well, or maybe the relative position or how much that node is moved relative to its absolute or relative position during storing. Unfortunately, this requires that a record is enough to store both the positional information and the number of idle integers practiced by that node. However, as explained earlier, further associations can be created using the idle integers that were already practiced by manipulating either their position or value or both. Hence, if the record is enough, it can store both the positional information and the number of idle integers. If not, an idle integer can be associated accompanying the node to supply additional space for it for the positional information.

Finally, the sorted permutation of the practiced interval is constructed in the list space, using the stored information in the short-term memory. This is the {\em retrieval} phase of associative sort that depends on the information encoded into the record of a node. If the record is enough, it stores both the position of the node and the number of idle integers. If not, an associated idle integer accompanying the node stores the position of the node while the record holds the number of idle integers. The positional information cues the recall of the integer using the inverse hash function. This is ``integer retrieval'' from imaginary subpace. Hence, the retrieved integer can be copied on the list space as many as it occurrs.

Hence, moving through nodes that represent the start and end of practiced integers as well as retaining their relative associations with each other even when their positions are altered by cuing allow the order of integers to be constructed in linear time in-place.

From complexity point of view, associative sort shows similar characteristics with bucket sort~\cite{mahmoud:2000,Cormen} and distribution counting sort~\cite{Seward,Feurzig}. It sorts $n$ integers $S[0\ldots n-1]$ each in the range $[0,m-1]$ using $\mathcal{O}(1)$ extra space in $\mathcal{O}(n+m)$ time for the worst, $\mathcal{O}(m)$ time for the average (uniformly distributed integers) and $\mathcal{O}(n)$ time for the best case. The ratio $\frac{m}{n}$ defines the efficiency (time-space trade-offs) of the algorithm letting very larges lists to be sorted in-place. 

\subsection{In-place Associative Permutation Sort}

In this study, associative sorting technique is revised combining the ideas behind the ordinal theories of Shiffrin and Cook~\cite{Shiffrin} and the original perturbation model of Estes~\cite{Estes}. In perturbation model of Estes, the order is inherent in the cyclic reactivation of elements. When both theories are combined, in-place associative permutation sort technique is obtained where the order of the practiced interval is inherent in the cyclic reactivation of the elements of the list. Hence, reactivating (stimulating) the list with a special form of cycle leader permutation result in a temporal state that can be either stored in short (or long) term memory or restored into sorted permutation of the practiced interval all in linear time using $\mathcal{O}(1)$ amount of memory with three phases: (i) {\em practicing}, (ii) {\em permutation (stimulation, reactivation)} and (iii) {\em restoring}. 

Practical comparisons for lists up to one million integer keys all in the range $[0,n-1]$ on a Pentium machine with radix sort and bucket sort indicate that associative permutation sort is slower roughly $2$ times than radix sort and slower roughly $3$ times than bucket sort. On the other hand, it is faster than quick sort for the same lists roughly $1.5$ times.
 
Although its time complexity is similar to that of in-place associative sort~\cite{ecetin} and practically slower, in-place associative permutation sort is proposed for {\em integer key sorting} problem.  Hence, it sorts $n$ elements $S[0\ldots n-1]$ each have an integer key in the range $[0,m-1]$ with $m>n$ using $\mathcal{O}(1)$ extra space in $\mathcal{O}(n+m)$ time for the worst, $\mathcal{O}(m)$ time for the average (uniformly distributed keys) and $\mathcal{O}(n)$ time for the best case.

%%%%%%%%%%%%%%%%%%%%%%%%%%%%%%%%%%%%%%%%%%%%%%%%%%%%%%%%%%%%%%%%%%%%%%%%%%%%%%%%%%%%%%%%%%%%%%%%%%%%%%%%%
%%%%%%%%%%%%%%%%%%%%%%%%%%%%%%%%%%%%%%%%%%%%%%%%%%%%%%%%%%%%%%%%%%%%%%%%%%%%%%%%%%%%%%%%%%%%%%%%%%%%%%%%%
\section{Definitions}\label{sec:pre}
Given a {\em list} $S$ of $n$ {\em elements}, $S[0], S[1],\ldots , S[n-1]$ each have an integer {\em key}, the problem is to sort the elements of the list according to their integer keys. To prevent repeating statements like ``key of the element $S[i]$'', $S[i]$ is used to refer the key. The notations used throughout the study are: 
\begin{enumerate} [label=({\roman{*}}), nosep]
\item Universe of integer keys is assumed $\mathbb{U} = [ 0 \ldots 2^{w}-1]$ where $w$ is the fixed word length.

\item Maximum and minimum keys of a list are, $\max (S) = \max(a \vert a \in S)$ and $\min (S) = \min(a \vert a \in S)$, respectively. Hence, range of the keys is, $m = \max (S) - \min (S) + 1$.

\item The notation $B \subset A$ is used to indicated that $B$ is a proper subset of $A$.

\item For two lists $S_{1}$ and $S_{2}$, $\max (S_{1}) < \min (S_{2})$ implies $S_{1} < S_{2}$.

\end{enumerate}

\begin{description}[leftmargin = 0pt]

\item[{\bf Universe of Keys.}] When a key is first practiced, a node is created releasing $w$ bits of the key free. One bit is used to tag the word as a node. Hence, it is reasonable to doubt that the tag bit limits the universe of keys because all the keys should be untagged and in the range $[0,2^{w-1}-1]$ before being practiced. But, we can,
\begin{enumerate}[label=(\roman{*}), itemindent = * , nosep]
\item partition $S$ into $2$ disjoint sublists $S_1 < 2^{w-1} \le S_2$ in $\mathcal{O}(n)$ time with well known in-place partitioning algorithms or stably with~\cite{Katajainen},
\item shift all the keys of $S_2$ by $-2^{w-1}$, sort $S_1$ and $S_2$ with associative permutation sort technique and shift $S_2$ by $2^{w-1}$.
\end{enumerate}
There are other methods to overcome this problem. For instance, 
\begin{enumerate}[label=(\roman{*}), itemindent = * , nosep]
\item sort the sublist $S[0\ldots (n/ \log n)-1]$ using the optimal in-place merge sort~\cite{Salowe},
\item compress $S[0\ldots (n/ \log n)-1]$ by Lemma~1 of~\cite{Franceschini_1} generating $\Omega(n)$ free bits,
\item sort $S[(n/ \log n)\ldots n-1]$ with in-place associative permutation sort technique using $\Omega(n)$ free bits as tag bits,
\item uncompress $S[0\ldots (n/ \log n)-1]$ and merge the two sorted sublists in-place in linear time by~\cite{Salowe}.
\end{enumerate}

\item [{\bf Number of Keys.}] If practicing a distinct key lets us to use $w-1$ bits to practice other occurrences of that particular key, we have $w-1$ free bits by which we can count up to $2^{w-1}$ occurrences including the first key that created the node. Hence, it is reasonable to doubt again that there is another restriction on the size of the lists, i.e., $n \le 2^{w-1}$. But a list can be divided into two parts in $\mathcal{O}(1)$ time and those parts can be merged in-place in linear time by~\cite{Salowe} after sorted associatively.

\end{description}

Hence, for the sake of simplicity, it will be assumed that $n \le 2^{w-1}$ and all the keys are in the range $[0,2^{w-1}-1]$ throughout the study.

%%%%%%%%%%%%%%%%%%%%%%%%%%%%%%%%%%%%%%%%%%%%%%%%%%%%%%%%%%%%%%%%%%%%%%%%%%%%%%%%%%%%%%%%%%%%%%%%%%%%%%%%%
%%%%%%%%%%%%%%%%%%%%%%%%%%%%%%%%%%%%%%%%%%%%%%%%%%%%%%%%%%%%%%%%%%%%%%%%%%%%%%%%%%%%%%%%%%%%%%%%%%%%%%%%%

\section{Associative Permutation Sort} \label{sec:basics}

Given $n$ {\em distinct} integer keys $S[0\ldots n-1]$ each in the range $[u, v]$, if $m=n$, the sorted permutation of the list can be represented with two parameters ($2 \log \mathbb{U}$ bits) one of which is the initial address of the sequential memory separated for the list (accessed by $S[0]$) in the RAM and the other is the $\delta=u$. The $i$th key of the sorted list can be calculated by $S[i]=i+\delta$ and the actual value at $i$th location is meaningless for this calculation. Hence, if $S$ is the sorted permutation, then there is a bijective relation between each key and its position, i.e., $i=S[i]-\delta$. From contradiction, if $S$ is not the sorted permutation, $i \ne S[i]-\delta$ implies that the key $S[i]$ is not at its exact location. Its exact location can be calculated with $j=S[i]-\delta$. Therefore, this monotone bijective hash function that maps the keys to $j \in [0,n-1]$ can sort the list in $\mathcal{O}(n)$ time using $\mathcal{O}(1)$ constant space. This is cycle leader permutation where $S$ is re-arranged by following the cycles of a permutation $\pi$. First $S[0]$ is sent to its final position $\pi(0)$ (calculated by $j=S[i]-\delta$). Then the element that was in $\pi(0)$ is sent to its final position $\pi(\pi(0))$. The process proceeds in this way until the cycle is closed, that is until the key to position $0$ is found which means that the association $0 = S[0] - \delta$ is constructed between the first key and its position. Then the iterator is increased to continue with the key of $S[1]$. At the end, when all the cycles of $S[i]$ for $i=0,1..,n-1$ are processed, all the keys are moved to their exact position and the association $i = S[i] - \delta$ is constructed between the keys and their position, i.e., the sorted permutation of the list is obtained. 

Associative permutation sort is based on this very simple idea: if the keys of a predefined interval are all distinct, then the list can be reactivated (stimulated) starting a special form of cycle leader permutation which rearranges the practiced interval into a state that can be simply restored into sorted permutation. 

An ILS is defined as $Im[0...n-1]$ over $S[0...n-1]$ if the interval of range of keys that it spans is exactly equal to $n$. Hence, an ILS $Im[0 \ldots n-1]$ can span the keys that are in $[\delta,\delta+n-1]$ where $\delta=\min(S)$. With this information, we can state that,
\begin{lem}\label{lem:sorting_seq_of_distinct}
Given $n$ integer keys $S[0\ldots n-1]$, the keys that are in the range $[\delta, \delta+n-1]$ where $\delta=\min(S)$ can be sorted at the beginning of the list in $\mathcal{O}(n)$ time using $\mathcal{O}(1)$ constant space.
\end{lem}

\begin{proof}

Given $n$ distinct integer keys $S[0...n-1]$ each in the range $[u, v]$, it is not possible to construct a monotone bijective hash function ({\em minimal monotone perfect hash function})  that maps all the keys of the list into $j \in [0,n-1]$ without additional storage space~\cite{Belazzougui}. However, a monotone bijective hash function can be constructed as a partial function~\cite{rosen:discrete_math_handbook} that assigns each key of $S_1 \subset S$ in the range $[\delta,\delta+n-1]$ with $\delta=\min(S)$ to exactly one element in $j \in [0,n-1]$ ignoring the keys of $S_2 = S - S_1$. The partial monotone bijective hash function of this form is,
\begin{equation}\label{eqn:hash_func_mul}
\begin{split}
j=S[i]- \delta  \quad \text{if} \quad S[i] - \delta  < n
\end{split}
\end{equation}

With this definition, the proof has three basic steps of associative permutation sort: 
\begin{enumerate}[label=(\roman{*})]
\item Practice all the keys in the range $[\delta, \delta+n-1]$. If $n_d$ nodes are created that practiced $n_c$ idle keys, then all the keys in the practiced interval become distinct and consecutive in the range $[0 \ldots n_d+n_c-1]$ after practicing phase ignoring the tag bits.
\item Permute (stimulate, reactivate) the list to rearrange the practiced interval at the beginning of the list.
\item Restore the sorted permutation of the practiced interval. This phase is only to bring the keys to their original values. The elements of the list (satellite data) are already sorted before this phase.
\end{enumerate}
\end{proof}

%%%%%%%%%%%%%%%%%%%%%%%%%%%%%%%%%%%%%%%%%%%%%%%%%%%%%%%%%%%%%%%%
\subsection{Practicing Phase}\label{sec:practicing}

\begin{enumerate}[label=\bf{Algorithm \Alph{*}.}, ref=Algorithm \Alph{*}, leftmargin=0pt, itemindent=*, start=1] 
\item \label{algorithm:counting} Practice all the keys in the range $[\delta,\delta+n-1]$ by mapping them into ILS $Im[0...n-1]$ over $S[0...n-1]$ using Eqn.~\ref{eqn:hash_func_mul}. It is assumed that the minimum of the list $\delta=\min(S)$ is known.
\end{enumerate}
\begin{enumerate}[label=\bf{A\arabic{*}.}, ref=A\arabic{*}, itemindent=*]
\item initialize $i = 0$;\label{algo1:item1}
\item if MSB of $S[i]$ is $1$, then $S[i]$ is a node. Hence, increase $i$ and repeat this step;\label{algo1:item2}
\item if $S[i] - \delta  \ge n$, then $S[i]$ is a key of $S_2$ which is out of the interval that is practiced. Hence, increase $n_d'$ that counts the number of keys of $S_2$, update $\delta'=min(\delta', S[i])$, increase $i$ and goto to step \ref{algo1:item2};\label{algo1:item2_3}
\item otherwise, $S[i]$ is a key of $S_1$ that is to be practiced. Hence, calculate $j = S[i] - \delta$ (Eqn.~\ref{eqn:hash_func_mul});\label{algo1:item3}
\item if MSB of $S[j]$ is $0$, then $S[i]$ is the first key that will create the node at $j$, hence move $S[j]$ to $S[i]$, clear $S[j]$ and set its MSB to $1$. If $j \le i$ increase $i$. Increase $n_d$ that counts the number of distinct keys and hence the nodes, and goto step \ref{algo1:item2}; \label{algo1:item4}
\item otherwise, $S[j]$ is a node that has already been created by another occurrence of $S[i]$. Hence, clear MSB of $S[j]$, increase $S[j]$ and set its MSB back to $1$. Increase $i$ and $n_c$ that counts the number of total idle keys over all distinct keys and goto step \ref{algo1:item2};\label{algo1:item5}
\end{enumerate}
At the end of practicing, each record of a node keeps the number of idle keys practiced by that node. Hence, the record of a node at $i$ stores one less (the node itself represents the first occurrence that creates the node) the number of occurrences of a key equal to $i+\delta$. In total, $n_d$ nodes are created that practice $n_c$ idle keys.

The next step of practicing phase is the accumulation.
\begin{enumerate}[label=\bf{Algorithm \Alph{*}.}, ref=Algorithm \Alph{*}, leftmargin=0pt, itemindent=*, start=2] 
\item \label{algorithm:summing} Sum all the records of the nodes from left to right.
\end{enumerate}
\begin{enumerate}[label=\bf{B\arabic{*}.}, ref=B\arabic{*}, itemindent=*]
\item initialize $i = 0$ and $j=0$;\label{algo2:item1}
\item if MSB of $S[i]$ is $1$, then $S[i]$ is a node. Hence, clear MSB of $S[i]$, set $S[i] = S[i] + j$ and $j = S[i]$, set MSB of $S[i]$ back, increase $i$ and repeat this step;\label{algo2:item2}
\item otherwise, increase $i$ and goto step \ref{algo2:item2}; 
\end{enumerate}
At the end, each record of a particular node keeps the exact position of the last idle key practiced by that node.

An ILS can create other subspaces and associations using the idle keys that were already practiced by manipulating either their position or value or both. Hence, it is logical to use the nodes of ILS as discrete hash functions that define the values of idle keys when they are re-practiced using the same monotone bijective hash function (Eqn.~\ref{eqn:hash_func_mul}). This is the {\em re-practicing} step that makes all the idle keys and the nodes distinct (ignoring the tag bits),

\begin{enumerate}[label=\bf{Algorithm \Alph{*}.}, ref=Algorithm \Alph{*}, leftmargin=0pt, itemindent=*, start=3] 
\item \label{algorithm:differentiation} Re-practice all the idle keys by mapping them again to ILS $Im[0...n-1]$ over $S[0...n-1]$ using the same hash function (Eqn.~\ref{eqn:hash_func_mul}). When an idle key is remapped to its node, it will obtain its exact position (ticket) from its node which will make it distinct and in the range $[0, n-1]$. The record of the node will be decreased by one for each re-practiced idle key. Hence, when all the idle keys of a node are re-practiced, the node will become distinct as well, in the range $[0, n-1]$ storing its exact position when its tag bit (MSB) is ignored. It should be noted that, if \ref{algorithm:counting} was stable, the list should be processed from right to left for stability as below.

\end{enumerate}
\begin{enumerate}[label=\bf{C\arabic{*}.}, ref=C\arabic{*}, itemindent=*]
\item initialize $i = n-1$;\label{algo3:item1}
\item if MSB of $S[i]$ is $1$, then $S[i]$ is a node. Hence, decrease $i$ and repeat this step;\label{algo3:item2}
\item if $S[i] - \delta  \ge n$, then $S[i]$ is a key of $S_2$ that is out of practiced interval. Hence, decrease $i$ and goto to step \ref{algo3:item2};\label{algo3:item2_1}
\item otherwise, $S[i]$ is an idle key. Hence, calculate $j = S[i] - \delta$ (Eqn.~\ref{eqn:hash_func_mul}), clear MSB of $S[j]$, copy $S[j]$ over $S[i]$, decrease $S[j]$ by one and set its MSB back to $1$, decrease $i$ and goto step \ref{algo3:item2};\label{algo3:item3}
\end{enumerate}
At the end, when the tag bits are ignored, all the keys and nodes of the practiced interval became distinct, in the range $[0, n-1]$ and their distinct values are equal to their exact positions in the sorted permutation.

%%%%%%%%%%%%%%%%%%%%%%%%%%%%%%%%%%%%%%%%%%%%%%%%%%%%%%%%%%%%%%%%
\subsection{Permutation Phase}\label{sec:stimulation}

In the permutation phase, all the elements of the list are reactivated (stimulated) starting a special form of cycle leader permutation that leads to reordering of the practiced interval in short-term memory ($S[0 \ldots n_d+n_c-1]$) which can be simply restored into sorted permutation of the original keys. It is not possible to use a simple cycle leader permutation because there are nodes and if their position change, the association between ILS and the list space is broken. However, a node has a record of $w-1$ bits which at the moment keeps the node's exact position in the sorted permutation. Hence, when a node is moved to its exact position, its former position can be overwritten into its record as the cue that can be used to recall the key using the inverse hash function, i.e., retrieve it from ILS. But, how an algorithm can distinguish the nodes that are already moved, from the nodes that are not moved yet in such a case? The idle keys and the keys that are out of the practiced interval are the solution to this problem. If a node is at wrong position, then it is evident that either an idle key or a key that is out of the practiced interval is available which will address that position. Hence, a special outer cycle leader permutation that reactivates only the idle keys and the keys that are out of the practiced interval will ensure that the corresponding one will be moved to the actual position of the node giving a chance to start an inner cycle leader permutation that reactivates only the nodes and ensures that the node will be moved to its exact position keeping its former position in its record as the cue. Once a node is moved to its exact position, there can not be any other outer cycle leader which will address that particular position. If another node is available where that particular node is moved, then the inner cycle leader permutation can continue with that node. However, if an idle key or a key that is out of the practiced interval is encountered, then the inner cycle leader permutation ends and the outer cycle leader permutation continues.

\begin{enumerate}[label=\bf{Algorithm \Alph{*}.}, ref=Algorithm \Alph{*}, leftmargin=0pt, itemindent=*, start=4] 
\item \label{algorithm:stimulation} Permute (stimulate, reactivate) the list from left to right to move all the idle keys and the nodes to their exact positions in the short-term memory ($S[0 \ldots n_d+n_c-1]$) from where one can restore the original keys. For this purpose, start an outer cycle leader permutation that reactivates only the idle keys and the keys that are out of the practiced interval. When a key is found that is out of the practiced interval, move it to $S[k]$ starting with $k=n_d+n_c$, and increase $k$ every time a key out of the practiced interval is moved and continue with the key that was at $k$ before. If an idle key is found, implicitly practice it, i.e., move it to $S[j]$ where $j = S[i]$. If an idle key or a key that is out of the practiced interval is moved to a position where a node is there, start an inner cycle leader permutation that reactivates only the nodes until a new idle key or a key that is out of the practiced interval is encountered. Repeat this until all the idle keys of the list are reactivated and all the keys that are out of the practiced interval are moved to $S[n_d+n_c \ldots n-1]$. At the end, a list is obtained where all the distinct idle keys are in their exact position in the sorted permutation, i.e., they are implicitly practiced as in ~\cite{ecetin} and an association is created between each idle key and its position with the monotone bijective hash function $i = S[i]$, as well as the nodes are in their exact positions and each node precedes all the idle keys that it has practiced, from where one can obtain the exact values of those idle keys by retrieving the key back from ILS through that node using its record as the cue.
\end{enumerate}
\begin{enumerate}[label=\bf{D\arabic{*}.}, ref=D\arabic{*}, itemindent=*]
\item initialize $i = 0$, $k=n_d+n_c$;\label{algo4:item1}
\item if MSB of $S[i]$ is $1$, then $S[i]$ is a node. Hence, increase $i$ and repeat this step;\label{algo4:item2}
\item if $i=S[i]$, then $S[i]$ is an idle key that has already been practiced implicitly. Hence, increase $i$ and goto step~\ref{algo4:item2};\label{algo4:item2_2}

\item $S[i]$ is either an idle key not practiced yet or a key that is out of the practiced interval. Hence;\label{algo4:item3}

\begin{enumerate}[label=(\roman{*})]
\item if $S[i]$ is an idle key, i.e., in the range $[0,n-1]$, then swap $S[i]$ with $S[j]$ where $j=S[i]$;\label{algo4:item3_1}
\item otherwise, $S[i]$ is a key that is out of the practiced interval. Hence, swap $S[i]$ with $S[k]$, set $j=k$ and increase $k$, \label{algo4:item3_2}
\end{enumerate}

\item if MSB of $S[i]$ is $0$, then $S[i]$  is either an idle key or a key that is out of the practiced interval again. Hence, goto step \ref{algo4:item2_2};\label{algo4:item4}
\item otherwise, $S[i]$ is a node. Hence, start an inner cycle leader permutation. Clear MSB of $S[i]$, swap $S[i]$ with $S[p]$ where $p=S[i]$, encode former position $j$ (comes from step \ref{algo4:item3}) of the node into its record (least significant $w-1$ bits of $S[p]$) and set MSB of $S[p]$ back to $1$;\label{algo4:item5}
\item if MSB of $S[i]$ is $1$, then it is a node again. Hence, continue the inner cycle leader permutation, i.e., set $j = p$ and goto step~\ref{algo4:item5};\label{algo4:item7}
\item otherwise, $S[i]$ is either an idle key or a key that is out of the practiced interval. Hence, finish inner cycle leader permutation and goto step~\ref{algo4:item2_2};\label{algo4:item6}
\end{enumerate}
At the end, a list is obtained where,
\begin{enumerate}[label=(\roman{*}) ]
\item all the distinct idle keys are in their exact position in the sorted permutation, i.e., they are implicitly practiced as in ~\cite{ecetin} and an association is created between each idle key and its position with the monotone bijective hash function $i = S[i]$,
\item the nodes are in their exact position and each node precedes its idle keys, from where one can obtain the exact values of those idle keys by retrieving the key back from ILS through that node using its record as the cue,
\item $n_d'$ unpracticed keys are located disorderly at $S[n_d+n_c \ldots n-1]$.

\end{enumerate}

\subsection{Restoring Phase}\label{sec:restore}

With a final scan of the short-term memory ($S[0 \ldots n_d+n_c-1]$), one can obtain the exact values of the practiced keys at $S[0 \ldots n_d+n_c-1]$. Hence, this phase is only to bring the keys to their original values. The elements of the list (satellite data) are already sorted before this phase.
\begin{enumerate}[label=\bf{Algorithm \Alph{*}.}, ref=Algorithm \Alph{*}, leftmargin=0pt, itemindent=*, start=5] 
\item \label{algorithm:construction} Restore the sorted permutation of the practiced interval.
\end{enumerate}
\begin{enumerate}[label=\bf{E\arabic{*}.}, ref=E\arabic{*}, itemindent=*]
\item initialize $i = 0$;\label{algo5:item1}
\item if MSB of $S[i]$ is $1$, then $S[i]$ is a node. Hence, decode the absolute position $j$ of the node from its record ($w-1$ bits of $S[i]$) and cue the recall of the key using the inverse hash function $k=j+\delta$ and retrieve the key from ILS by $S[i] = k$. \label{algo5:item2}
\item increase $i$. If $S[i]$ is in the range $[0,n-1]$, then it is an idle key of the preceding node. Hence, copy the exact value of the key by $S[i]=k$ and repeat this step;\label{algo5:item3}
\item if MSB of $S[i]$ is $1$, then it is a new node. Hence, goto step \ref{algo5:item2};\label{algo5:item4}
\item otherwise, $S[i]$ is a key that is out of the practiced interval. Hence, exit. 
\end{enumerate}

\subsection{Binding the Loop}\label{sec:overall}

After restoring, $n_d+n_c$ practiced keys are sorted at the beginning of the list while unpracticed $n_d'$ keys of $S_2$ are distributed disorderly at $S[n_d+n_c \ldots n-1]$. Hence, the structure of the sequential version becomes,
\begin{enumerate}[label=\bf{Algorithm \Alph{*}.}, ref=Algorithm \Alph{*}, leftmargin=0pt, itemindent=*, start=4] 
\item \label{algorithm:es_fgd_iter_mul} In each iteration, construct sorted permutation of $n_d+n_c$ keys of $S_1$ at the beginning of the list.
\end{enumerate}
\begin{enumerate}[label=\bf{D\arabic{*}.}, ref=D\arabic{*}, itemindent=* , nosep]
\item find $\min(S)$ and $\max(S)$;\label{algo20:item1}
\item initialize $\delta = \min(S)$, $\delta' = \max(S)$ and reset counters;\label{algo20:item2}
\item practice all the keys in the interval $[\delta,\delta+n-1]$ using \ref{algorithm:counting} to \ref{algorithm:differentiation};\label{algo20:item3}
\item permute the list using \ref{algorithm:stimulation};\label{algo20:item4}
\item restore the sorted permutation of the practiced interval using \ref{algorithm:construction};\label{algo20:item6}
\item if $n_d'=0$ exit. Otherwise set $S=S[n_d+n_c \ldots n-1]$, $n = n_d'$, $\delta = \delta'$, $\delta' = \max(S)$, reset counters and goto step \ref{algo20:item3}.\label{algo20:item7}
\end{enumerate}

% \begin{rem}
% $\min(S)$ and $\max(S)$ need not be found in step \ref{algo20:item1}. Instead, if $\delta = 0$ and $\delta' = \max(\mathbb{U})$ the algorithm sorts the keys in the range $[0,n-1]$ during the first iteration (or recursion). However, if there is not any key in this interval, \ref{algorithm:counting} finds $\delta'=\min(S)$ in step \ref{algo1:item2_3} in $\mathcal{O}(n)$ time, and continues with the keys in $[\delta', \delta'+n-1]$.
% \end{rem}

\begin{rem}
Associative permutation sort technique is on-line in the sense that after each step \ref{algo20:item6}, $n_d+n_c$ keys are added to the sorted permutation at the beginning of the list and ready to be used.
\end{rem}

\begin{description}[leftmargin = 0pt]
\item[{\bf Practical comparisons}] for lists up to one million integer keys all in the range $[0,n-1]$ on a Pentium machine with radix sort and bucket sort indicate that associative permutation sort is slower roughly $2$ times than radix sort and slower roughly $3$ times than bucket sort. On the other hand, it is faster than quick sort for the same lists roughly $1.5$ times.

\end{description}

%**************************************************************************************
\begin{description}[leftmargin = 0pt]

\item [{\bf Complexity}] of the algorithm depends on the range and the number of the keys. In each iteration (or recursion) the algorithm is capable of sorting keys that satisfy $S[i] - \delta < n$. Hence, given $n$ integer keys $S[0 \ldots n-1]$ each in the range $[0,n-1]$, the complexity is $T(n) = \mathcal{O}(n)$. 

\item[{\bf Best Case Complexity}] Given $n$ integer keys $S[0 \ldots n-1]$ each in the range $[0,\mathbb{U}]$, if $n-1$ keys satisfy $S[i] - \delta < n$, then these keys are sorted in $\mathcal{O}(n)$ time. In the next step, there is $n'=1$ key left which implies that the sorting is finished. As a result, time complexity of the algorithm is lower bounded by $\Omega(n)$ in the best case.

\item [{\bf Worst Case Complexity}] Given $n$ integer keys $S[0 \ldots n-1]$ each in the range $[u,v]$ and $m=v-u+1=\beta n$, if there is only $1$ key available in the practiced interval at each iteration, in any $j$th step, the only key $s$ that will be sorted satisfies $s < {jn-(j-1)}$, which implies that the last alone key satisfies $s < {jn-(j-1)} < \beta n$ from where we obtain $j < \frac{\beta n-1 }{n-1}$. In this case, the time complexity of the algorithm is, 
\begin{equation}
\mathcal{O}(n) + \mathcal{O}(n-1) + \dotsc + \mathcal{O}(n-j) = (j+1) \mathcal{O}(n) -\mathcal{O}(j^2) < (\beta+1) \mathcal{O}(n)
\end{equation}
Therefore, the time complexity of the algorithm in worst case is $(\beta+1) \mathcal{O}(n) = \mathcal{O}(m+n)$.

\item[{\bf Average Case Complexity.}] Given $n$ integers $S[0 \ldots n-1]$, if $m = \beta n$ and the integers are uniformly distributed, this means that $\frac{n}{\beta}$ integers satisfy $S[i] < n$. Therefore, the algorithm is capable of sorting $ \frac{n}{\beta}$ integers in $\mathcal{O}(n)$ time during first pass. This will continue until all the integers are sorted. The sum of sorted integers in each iteration can be represented with the series,
\begin{equation} \label{eqn:series_sorted_case3}
\frac{n}{\beta}+\frac{n(\beta-1)}{\beta^2}+\dotsc+\frac{n(\beta-1)^{k-1}}{\beta^{k}}+\dotsc
\end{equation}

It is reasonable to think that the sorting ends when one term is left which means the sum of $k$ terms of this series is equal to $n-1$, from where we can calculate the number of iteration or dept of recursion $k$ which is valid when $\beta > 1$ by,
\begin{equation} \label{eqn:series_sorted_case3_4}
\frac{1}{n} = \frac{(\beta-1)^{k-1}}{\beta^{k}}
\end{equation}
It is seen from Eqn.~\ref{eqn:series_sorted_case3_4} that when $m = 2n$, i.e., $\beta=2$, number of iteration or dept of recursion becomes $k=\log{n}$ and the complexity is the recursion $T(n) = T(\frac{n}{2}) + \mathcal{O}(n)$ yielding $T(n) = \mathcal{O}(n)$. It is known that each step takes $\mathcal{O}(n)$ time. Therefore, the time complexity of the algorithm is,
\begin{equation}\label{eqn:series_complexity_case3}
\begin{split}
\mathcal{O}(n)+\mathcal{O}\bigl(\frac{n(\beta-1)}{\beta}\bigr) +\dotsc +\mathcal{O}\bigl( \frac{n(\beta-1)^{k-1}}{\beta^{k-1}} \bigr)
\end{split}
\end{equation}
from where we can obtain by defining $x= \frac{(\beta-1)}{\beta}$,
\begin{equation}\label{eqn:ud_6}
\mathcal{O}(n) \bigl( 1 +  x + x^2 + x^3 + \cdots + x^{k-1} \bigr) = \mathcal{O}(n) (\frac{1}{1-x} - \frac{x^{k-1}}{1-x}) < \beta \mathcal{O}(n)
\end{equation}
which means that the algorithm is upper bounded by $ \beta \mathcal{O}(n)$ or $\mathcal{O}(m)$ in the average case.

\end{description}

%**************************************************************************************

\section{Conclusions}
\label{chap:summaryandconclusion}
%****************************************************************************

In-place associative permutation sort technique is proposed which solves the main difficulties of distributive sorting algorithms by its inherent three basic steps namely (i) practicing, (ii) permutation and (iii) restoring. It is very simple and straightforward and around 30 lines of C code is enough. 

From time complexity point of view, associative permutation sort shows similar characteristics with bucket sort. It sorts the keys associatively in $\mathcal{O}(m)$ time for the average (uniformly distributed keys) and $\mathcal{O}(n)$ time for the best case. Although its worst case time complexity is $\mathcal{O}(n+m)$, it is upper bounded by $\mathcal{O}(n^2)$ for the lists where $m>n^2$. On the other hand, it requires only $\mathcal{O}(1)$ additional space, making it time-space efficient compared to bucket sort. The ratio $\frac{m}{n}$ defines its efficiency (time-space trade-offs) letting very large lists to be sorted in-place. Furthermore, the dependency of the efficiency on the distribution of the keys is $\mathcal{O}(n)$ which means it replaces all the methods based on address calculation, that are known to be very efficient when the keys have known (usually uniform) distribution and require additional space more or less proportional to $n$. Hence, in-place associative permutation sort asymptotically outperforms all content based sorting algorithms when $\frac{m}{n}<c$ and $c$ is the efficiency constant defined by the other sorting algorithms regardless of how large is the list.

The technique seems to be very flexible, efficient and applicable for other problems as well, such as hashing, searching, succinct data structures, gaining space, etc. 

The only drawback of the algorithm is that it is unstable. But, an imaginary subspace can create other subspaces and associations using the idle integers that were already practiced by manipulating either their position or value or both. Hence, different approaches can be developed to solve problems such as stability. 

\section{Discussion}
\label{chap:discussion}

Associative permutation sort first finds the minimum of the list and starts with the keys in $[\min(S), \min(S)+n-1]$. However, instead of starting with this interval, omitting the MSBs, if we consider a word as $w-1$ bits and the most significant $\lceil \log n \rceil$ bits of a word as the key and the remaining bits as the satellite information, the problem reduces to sorting $n$ integer keys $S[0 \ldots n-1]$ each in the range $[0,2^{\lceil \log n \rceil}-1]$. Since it is possible that $2^{\lceil \log n \rceil} -1 > n-1$, the keys in $[n, 2^{\lceil \log n \rceil}-1]$ become the keys that are out of the practiced interval.

%Another solution to this problem may seem to simply consider the most significant $\lfloor \log n \rfloor$ bits of a word as the key and the remaining bits as the satellite information which will reduce the problem to sorting $n$ integer keys $S[0 \ldots n-1]$ each in the range $[0,2^{\lfloor \log n \rfloor}-1]$. However, as there is always the risk that a key may occur more than $2^{\lfloor \log n \rfloor}-1$ times in a list of $n$ keys, it would not be possible to count that key in $\lfloor \log n \rfloor$ bits of corresponding node's record.

As a result, when the keys are sorted according to their most significant $\lceil \log n \rceil$ bits, {\em in-place associative most significant radix permutation sort} is obtained. After the list is sorted according to their most significant $\lceil \log n \rceil$ bits, the idle keys are grouped and each group is preceded by the corresponding node that has practiced them. Hence, each group can be sorted sequentially or recursively assuming the satellite information as the key. If itself is used, it becomes an algorithm based on {\em hash-and-conquer} paradigm in contrast to {\em divide-and-conquer}. However, size of subgroups decreases and it may not be efficient when the ratio of range of keys in each subgroup to size of that subgroup, i.e., $\frac{m}{n}$ increases. Hence, other strategies may need to be developed after the first pass.

%In worst case, half of the list is sorted depending on the difference $2^{\lceil \log n \rceil} - n$ which is upper bounded by $n$. Hence, time complexity of sorting the keys according to their most significant $\lceil \log n \rceil$ bits is the recursion $T(n)=T(n/2)+\mathcal{O}(n)$ yielding $T(n)=\mathcal{O}(n)$.

%\begin{acknowledgements}
%\end{acknowledgements}


%\bibliography{ecetin}{}


\end{document}